\documentclass{article}
\usepackage{amsmath,graphicx,mlspconf}   
\usepackage{url}            
\usepackage{booktabs}       
\usepackage{amsfonts}       
\usepackage{nicefrac}       
\usepackage{microtype}      
\usepackage{bm}
\usepackage{amsmath}
\usepackage{amsfonts}
\usepackage{color}
\usepackage[makeroom]{cancel}
\usepackage{cancel}
\usepackage{tabu}
\usepackage[normalem]{ulem} 
\usepackage{framed}
\usepackage{algorithm}
\usepackage{algorithmic}
\usepackage[draft,bookmarks=false]{hyperref} 
\usepackage{graphicx}
\usepackage{textcomp}
\usepackage{acronym}
\usepackage{tikz}
\usepackage{subfig}
\usepackage{multirow}
\usepackage{ctable} 
\usetikzlibrary{positioning}
\usepackage[textsize=tiny]{todonotes}
\acrodef{PSG}{Polysomnography}
\acrodef{HMM}{Hidden Markov Model}
\acrodef{HHMM}{Heterogeneous Hidden Markov Model}
\acrodef{OS}{Operative System}
\acrodef{SAR}{Sleep Activity Recognition}
\acrodef{HAR}{Human Activity Recognition}
\acrodef{EM}{Expectation-Maximization}
\acrodef{WASO}{Wake After Sleep Onset}
\acrodef{SE}{Sleep Efficiency}
\acrodef{SPT}{Sleep Period Time}
\acrodef{CM}{Circadian Midpoint}
\acrodef{SJ}{Social Jetlag}
\acrodef{BIC}{Bayesian Information Criteria}
\acrodef{AIC}{Akaike Information Criterion}
\acrodef{RMSE}{Root Mean Square Error}
\acrodef{MAE}{Mean Absolute Error}
\acrodef{SOL}{Sleep Onset Latency}
\acrodef{PSQI}{Pittsburgh Sleep Quality Index}
\acrodef{OSA}{Obstructive Sleep Apnea}
\acrodef{CPD}{Change Point Detection}
\acrodef{GMM}{Gaussian Mixture Model}

\def\0{{\mathbf 0}}

\title{Sleep Activity Recognition and Characterization from Multi-Source Passively Sensed Data}
\name{%
    María Martínez-García$^{*}$%
    \qquad Fernando Moreno-Pino$^{*}$%
    \qquad Pablo M. Olmos
    \qquad Antonio Art\'es-Rodr\'iguez%
    \thanks{$^{*}$ \normalsize{Martínez-García and F. Moreno-Pino are co-first authors with equal contributions.}}
}
\address{%
    Dept. of Signal Theory and Communications, Universidad Carlos III de Madrid, Spain \\%
    Gregorio Mara\~n\'on Health Research Institute, Spain%
}

\begin{document}

\maketitle

\begin{abstract}
Sleep constitutes a key indicator of human health, performance, and quality of life. Sleep deprivation has long been related to the onset, development, and worsening of several mental and metabolic disorders, constituting an essential marker for preventing, evaluating, and treating different health conditions. Sleep Activity Recognition methods can provide indicators to assess, monitor, and characterize subjects' sleep-wake cycles and detect behavioral changes. In this work, we propose a general method that continuously operates on passively sensed data from smartphones to characterize sleep and identify significant sleep episodes. Thanks to their ubiquity, these devices constitute an excellent alternative data source to profile subjects' biorhythms in a continuous, objective, and non-invasive manner, in contrast to traditional sleep assessment methods that usually rely on intrusive and subjective procedures. A Heterogeneous Hidden Markov Model is used to model a discrete latent variable process associated with the Sleep Activity Recognition task in a self-supervised way. We validate our results against sleep metrics reported by tested wearables, proving the effectiveness of the proposed approach and advocating its use to assess sleep without more reliable sources.
\end{abstract}

\section{Introduction}\label{introduction}

Sleep is a fundamental biological process that has long been considered a key determinant of human health and performance, as well as quality of life \cite{matsui2021association, gothe2020physical}. Poor sleep quality and sleep disturbances have been associated with the onset and development of several mental disorders \cite{scott2021sleep, freeman2020sleep, stowkowy2020sleep}, such as depression \cite{riemann2020sleep, nutt2022sleep, thase2022depression}, schizophrenia \cite{chung2018correlates, waite2020sleep}, bipolar disorder \cite{gold2016role, bradley2020association}, and suicidal risk increase  \cite{ geoffroy2021sleep, drapeau2019screening, porras2019contribution}. Moreover, sleep deprivation and sleep disorders, such as \ac{OSA}, have also been linked to the development and evolution of metabolic and cardiovascular diseases, including diabetes mellitus \cite{muraki2018sleep, perez2018sleep, song2019metabolic}, obesity \cite{mcmahon2019relationships, brum2020night}, and hypertension \cite{grandner2018sleep, tobaldini2019short, liu2019roles}. Thus, monitoring sleep can be vital to continuously assess the state of healthy and pathological subjects, acting as a biomarker for the prevention, evaluation, and treatment of patients.

Among traditional sleep assessment methods, \ac{PSG} stands as the most reliable technique, constituting the gold standard for studying and diagnosing sleep disorders. Nonetheless, \ac{PSG} requires medical assistance for its application, and sleep studies need to be recorded in an ambulatory setting \cite{smith2020actigraph, ibanez2018survey}. Furthermore, it is an expensive and obtrusive procedure that does not provide a continuous evaluation of patients' sleep. 
Sleep questionnaires such as the \ac{PSQI} \cite{buysse1989pittsburgh}, together with sleep diaries \cite{ibanez2018survey}, arise as a natural alternative when trying to monitor sleep evolution over time in a non-invasive manner. However, neither of these methods provides an objective sleep evaluation as they suffer from patients'
subjective insight, and they presume patients' ability and interest to fill in the required information \cite{hennig2017predicts, mallinson2019subjective, jeffs2019measuring}. 

\begin{figure*}[h!]
	\begin{center}
		\centerline{\includegraphics[width=1.7\columnwidth]{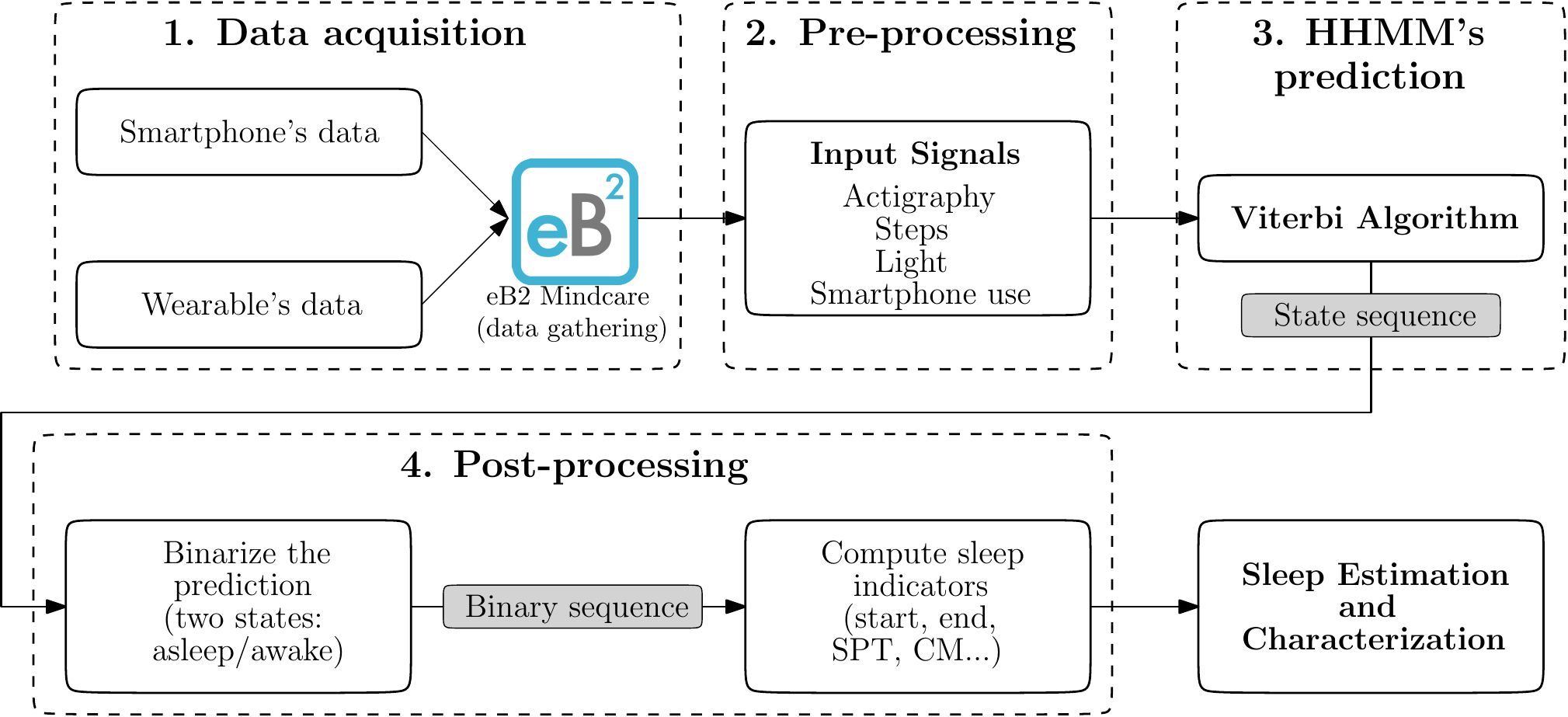}}
		\caption{Pipeline of the proposed Sleep Activity Recognition method.}
		\label{fig:scheme}
	\end{center}
\end{figure*}

In daily life, wearable devices constitute the best option to obtain objective, continuous sleep measurements in a non-invasive manner. Wearable devices feature multiple sensors, such as accelerometer and heart monitor, which allow to collect continuous daily information that can be used to provide sleep measurements. Although several studies have shown they are not sufficiently accurate to be used in clinical settings, mainly due to poor specificity and low accuracy, they can be relied on for detecting asleep-awake states \cite{moreno2019validation, berryhill2020effect, de2020sensors}. The main limitation of these wearable devices lies in the fact that the percentage of users who possess them is minimal, implying we cannot rely on wearables to assess large populations. 

Smartphones constitute an excellent data source for the problem of \ac{HAR}: their use requires no additional investment as most subjects already own smartphones, they are carried along most of the day by the users, and their ubiquity makes them the perfect passive assessment that does not interfere with subjects’ daily life. However, mobile health data collected this way usually entail several problems: missing data is quite common, signals are typically corrupted by noise, the differences in hardware and \acp{OS} across different models imply inter-user data heterogeneity, and main sleep trends must be detached from false positives while making the predictions.


In this work, we propose a \ac{SAR} method to estimate and characterize sleep. This method operates solely on objective, continuous, and passively sensed smartphone data. Similar state-of-the-art smartphone-based sleep detection methods \cite{saeb2017scalable, aledavood2018social, staples2017comparison} rely on location, WiFi, and microphone data to make predictions, which raise different users' privacy concerns. In this paper, we propose a method that can estimate sleep from non-intrusive sources, such as the light sensor, accelerometer data, smartphone usage indicators, and step count, handling heterogeneous and missing observations intrinsically. 

This work aims to present a general continuous data-based method to asses sleep using the most ubiquitous data source available: smartphones, therefore offering a feasible estimator when wearable or other devices are unavailable. Moreover, given the continuous nature of the proposed method, it could be used for patient behavior monitoring, as it does not interfere with subjects' daily life.

Although data gathered from wearable devices is not accurate enough to assess sleep in clinical settings and it cannot be relied on to diagnose sleep disorders, it can be used to estimate asleep-awake cycles and to depict sleep patterns in a continuous, objective, and non-invasive manner that is compatible with daily life. Furthermore, considering that a wide range of studies have reported that wearable devices show high reliability while detecting asleep-awake state changes in different populations, consistently reporting good estimators of the sleep activity \cite{de2018validation,chinoy2021performance}, we use them for validating purposes.

\section{Sleep Activity Recognition}\label{methods}

This section details our proposed solution for the \ac{SAR} problem. Fig. \ref{fig:scheme} shows a complete diagram of every step required to perform the predictions, from data collection to the model's inference. In the following subsections, we cover all parts of this pipeline, describing data collection and pre-processing model details and inference, and sleep characterization based on sleep indicators.

\subsection{Data Description}\label{data_description}

Smartphone data constitute a tool for subject monitoring due to their omnipresence, reporting objective measures of users' circadian rhythms and habits. The eB2 MindCare app \cite{eB2}, already used for data collection in different studies \cite{moreno2020passive, porras2020smartphone, sukei2021predicting, moreno2022heterogeneous}, was used to passively collect data from smartphones' sensors and patients' wearable devices such as Fitbit or Garmin. 

For this study, five passively sensed features from smartphone data were considered: actigraphy, light exposure, app usage (an indicator of smartphone usage, information about the app being used was not gathered), unlocks, and step count. Data's granularity was 10 minutes; hence 144 observations per day were available in the absence of missing data. We consider data from 14:00 of one day to 14:00 of the next day. Other studies have already used these data sources to model sleep-awake cycles \cite{saeb2017scalable, rahimi2021dpsleep, vaghela2022assessing}.

The available population consisted of a set of 41 non-pathological subjects. Participants were Master's and Ph.D. students from Universidad Carlos III de Madrid, Spain, with ages ranging from 22 to 30 years old. Data were collected anonymously under informed consent utilizing the eB2 MindCare app \cite{eB2}. The total population can be divided into two groups. The first one contains 17 of the 41 subjects, with 1523 days of data available. For this population, we have wearables data containing sleep metrics. Therefore, this subset is used for validation purposes exclusively, not taking part during the training process. The remaining 24 users have 687 days of complete data available without wearables' sleep metrics. Data from this subset is exclusively used for training purposes. 

\begin{figure*}[ht!]
\centering
    {\includegraphics[scale=0.5]{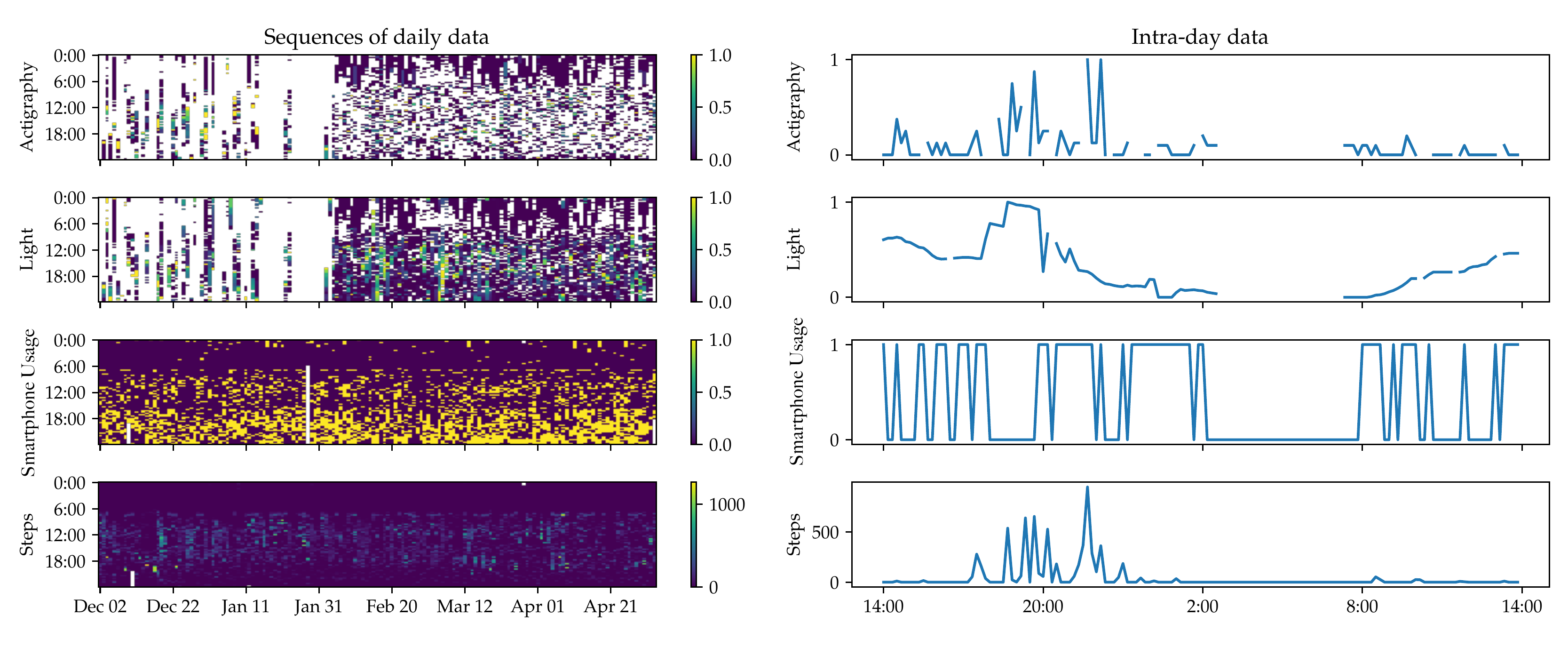}}
\caption{Subject's passively sensed data. Left figure shows five month of data, right figure displays intra-day features' values.}
\label{fig:dataset}
\end{figure*}

\subsection{Data Analysis}\label{data_analysis}

\subsubsection{Pre-processing}\label{data_pre-processing}
Since data have been collected from different sources and received in different formats, raw data present biases and anomalies, hence, noise. The presence of noise can impair the performance of machine learning methods, so data pre-processing is essential. Given the different nature of the collected signals, we define a specific pre-processing pipeline for each source. This processing is applied to each patient's 144-slots daily vector independently to tackle inter-user and intra-user variability.

\begin{itemize}
    \item \textbf{Actigraphy signal:} Samples deviated more than three times from the signal's standard deviation are identified as outliers and discarded. Then, we compute the signal mode to obtain the bias, which is different for each user, sensor, and device. This bias is removed, and samples with negative values are dismissed, as they are assumed erroneous. Finally, the resulting signal is normalized, taking values between zero and one.
    
    \item \textbf{Light signal:} We remove the outliers following the same criterion as in the actigraphy case. Then, we normalize the resulting signal between 0 and 1.
    
    \item \textbf{Step count:} As we are only interested in the presence or absence of activity at each time slot (we are not interested in the actual step count) we binarize this feature, setting it to one for samples with values larger than zero, thus obtaining a movement indicator.
    
    \item \textbf{App usage and unlocks:} These features provide complementary information about the same activity: smartphone use. Thus, we combine both signals to obtain a unique vector that indicates the use of the device during the day. Since both signals are already binary (app usage indicates whether any app was used at each time slot, and the unlocks feature indicates whether the smartphone was unlocked at each time slot), we combine them using the OR logical operator. 
\end{itemize}

Table \ref{tab:data} summarises the features generated by the pre-processing, while Fig. \ref{fig:dataset} shows the monthly and daily evolution of these features for an specific patient.

\begin{table}[h]
\fontsize{10.0}{14}\selectfont
\centering
\begin{tabular}{lc}
\hline
\textbf{Feature} & \textbf{Domain} \\\hline
Actigraphy          & $\mathbb{R}^{+}$ \\

Light               & $\mathbb{R}^{+}$ \\
Steps               & $\mathbb{Z}_{2}=\{0,1\}$\\
Smartphone usage    & $\mathbb{Z}_{2}=\{0,1\}$\\\hline
\end{tabular}
\caption{Resulting input feature vectors after pre-processing.}
\label{tab:data}
\end{table}

\subsubsection{Sleep Activity Recognition (SAR)}\label{sleep_activity_recognition}

To merge the heterogeneous input features into a \ac{SAR} indicator, capable of detecting different activity levels along the day and identifying major sleep episodes among them, we require a model capable of projecting the high dimensional input features into a lower dimensional space. To do so, latent variable models arise as a natural alternative \cite{bishop1998latent}. Considering the time dependence of our data, \acp{HMM} \cite{rabiner1989tutorial} constitute a natural way of modeling a discrete latent process associated with the \ac{SAR} task.

\acp{HMM} are generative models that assume the phenomena being modeled is a Markov process \cite{ethier2009markov} with associated hidden states. \acp{HMM}' objective is to learn about these hidden states sequences, denoted $S=\left\{s_{1}, s_{2}, \ldots, s_{T}: s_{t} \in 1, \ldots, I\right\}$, with $t=\left\{1, 2, \ldots, T\right\} \in  \mathbb{N}$, utilizing the generated observations, $\mathbf{Y}=\left\{\mathbf{y}_{1}, \mathbf{y}_{2}, \ldots, \mathbf{y}_{T}: \mathbf{y}_{t} \in \mathbb{R}^{M}\right\}$. They use an observation model $p(\mathbf{y}_t | s_t)$. Therefore, each observation $\mathbf{y}_t$ depends exclusively on its associated state $s_t$.

Here, we aim to use this hidden states sequence $S$ as the low dimensional representation of the input features. Therefore, each possible state of the HMM is associated with different activity levels, sleep episodes being one of them.

Nevertheless, standard \ac{HMM} implementations use Multinomial and Gaussian observation models, depending on the probability distribution chosen to model the emission probabilities $p(\mathbf{y}_t | s_t)$.
Due to the collected data's heterogeneity, neither option would fit our requirements. While some of the used features, like the actigraphy and light sensors, can be naturally modeled via Gaussian distributions, others, such as the smartphone usage and step count (binarized and converted into a movement indicator), are better modeled employing a Bernoulli distribution.

\subsection{\ac{HHMM}}\label{heterogeneous_hmm}

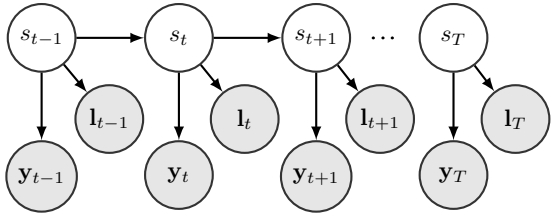
\begin{figure}[!htb]
\centering
\begin{tikzpicture}[scale=0.9, transform shape]
\tikzstyle{main}=[circle, minimum size = 10mm, thick, draw =black!80, node distance = 10mm]
\tikzstyle{connect}=[-latex, thick]
\tikzstyle{box}=[rectangle, draw=black!100]
  \node[main] (S1) [] {$s_{t-1}$};
  \node[main] (S2) [right=of S1] {$s_{t }$};
  \node[main] (S3) [right=of S2] {$s_{t+1}$};
  \node[main] (St) [right=of S3] {$s_{T}$};
  
  \node[main,fill=black!10] (O1) [below=of S1] {$\mathbf{y}_{t-1}$};
  \node[main,fill=black!10] (O2) [right=of O1,below=of S2] {$\mathbf{y}_{t}$};
  \node[main,fill=black!10] (O3) [right=of O2,below=of S3] {$\mathbf{y}_{t+1}$};
  \node[main,fill=black!10] (Ot) [right=of O3,below=of St] {$\mathbf{y}_{T}$};
  
  \node [main,fill=black!10] (L1) at (1,-1.2) {$\mathbf{l}_{t-1}$};
  \node [main,fill=black!10] (L2) at (3,-1.2) {$\mathbf{l}_{t}$};
  \node [main,fill=black!10] (L3) at (5,-1.2) {$\mathbf{l}_{t+1}$};
  \node [main,fill=black!10] (Lt) at (7,-1.2) {$\mathbf{l}_{T}$};

  \path (S3) -- node[auto=false]{\ldots} (St);
  \path (S1) edge [connect] (S2)
        (S2) edge [connect] (S3)
        (S3) -- node[auto=false]{\ldots} (St);

  \path (S1) edge [connect] (O1);
  \path (S2) edge [connect] (O2);
  \path (S3) edge [connect] (O3);
  \path (St) edge [connect] (Ot);
  \path (S1) edge [connect] (L1);
  \path (S2) edge [connect] (L2);
  \path (S3) edge [connect] (L3);
  \path (St) edge [connect] (Lt);

  \draw[dashed]  [below=of S1,above=of O1];
\end{tikzpicture}
\caption{Heterogeneous HMM architecture. Gray represents observed data.}
\label{hhmm}
\end{figure}

To manage the heterogeneity of our input sources, we developed a variation of ordinary \acp{HMM} which we refer to as Heterogeneous Hidden Markov Model (HHMM) \footnote{https://github.com/fmorenopino/HeterogeneousHMM}  \cite{moreno2022pyhhmm}. The proposed \ac{HHMM}, which could be employed for any \ac{HAR} task \cite{rios2020hidden, moreno2022heterogeneous}, can simultaneously manage categorical and continuous data thanks to its observation model, which is visible in Fig. \ref{hhmm}. This observation model is responsible for jointly modeling continuous and discrete observations, $\mathbf{y}_t$ and $\mathbf{l}_t$, respectively. In our case, continuous observations consist of light and actigraphy signals $\mathbf{y}_t = [actigraphy(t), light(t)]$, while discrete observations consist of step and smartphone usage signals $\mathbf{l}_t = [steps(t), smartphone(t)]$.

\acp{HHMM} can be fully characterized via the hidden states sequence, $S$; the continuous sequence observations, $\mathbf{Y}$; its associated continuous observations emission probabilities, $\mathbf{B}=\{b_{i}: p_{b_{i}}(\mathbf{y}_{t})=p(\mathbf{y}_{t} \mid s_{t}=i)\}$; the discrete sequence observations, $\mathbf{L}=\{\mathbf{l}_{1}, \mathbf{l}_{2}, \ldots, \mathbf{l}_{T}: \mathbf{l}_{t} \in 1, \ldots, J\}$; its associated discrete observations emission probabilities, $\mathbf{D}=\{d_{i m}: d_{i m}=P(l_{t}=m \mid s_{t}=i)\}$; the state transition probabilities, $\mathbf{A}=\{a_{i j}: a_{i j}=p(s_{t+1}=j \mid s_{t}=i)\}$; and the initial state probability distribution, $\pi=\{\pi_{i}: \pi_{i}=p(s_{1}=i)\}$.

\subsubsection{Parameter Inference}\label{parameter_inference}
    
\acp{HHMM}, as classic \acp{HMM}, presents three inference problems that must be solved to deliver a helpful application.

The first of these problems is related to obtaining the probability of the observed variables, $\mathbf{Y}$ and $\mathbf{L}$, given the model parameters, $\theta=\{\mathbf{A}, \mathbf{B}, \mathbf{D}, \mathbf{\pi}\}$, for the considered time-steps $t=\left\{1, 2, \ldots, T\right\}$, i.e., to calculate the probability $p(\mathbf{Y}, \mathbf{L} | \theta)$. 

The second problem consists of determining the optimal hidden state path that explains the observed data better. This can be achieved through the Forward-Backward algorithm, computing $p(s_t | \mathbf{y}_t, \mathbf{l}_t)$ each time-step, or through the Viterbi algorithm, which maximizes the probability of the hidden states sequences by considering all time-steps $t=\left\{1, 2, \ldots, T\right\}$, i.e., obtaining $p(S | \mathbf{y}_t, \mathbf{l}_t)$. 

Finally, the third problem consists of determining the optimal parameters $\theta$ that maximize the conditional probability $p(\mathbf{Y}, \mathbf{L} | \theta)$. The Baum-Welch algorithm, a special case of the \ac{EM} algorithm, can be used for obtaining this optimal parametrization. The joint distribution required for this third task is expressed in Eq. \ref{joint}. For further information regarding these three problems and their solution, we refer the readers to \cite{rabiner1989tutorial, moreno2022heterogeneous}.

\begin{equation}
\begin{aligned}
    p(S, Y, L) =& \prod_{n=1}^{N}\left(p\left(s_{1}^{n}\right) \prod_{t=2}^{T_{n}}  p\left(s_{t}^{n} | s_{t-1}^{n}\right)\right) \\
    &\left(\prod_{t=1}^{T_{n}} p\left(\mathbf{y}_{t}^{n} | s_{t}^{n}\right)\right)\left(\prod_{t=1}^{T_{n}} p\left(l_{t}^{n} | s_{t}^{n}\right)\right)
\end{aligned}
\label{joint}
\end{equation}

The proposed method can also be trained in a semi-supervised manner, fixing discrete observations' emission probabilities. When the model's parameters are trained in a semi-supervised way, discrete observations are referred to as labels.
This semi-supervised way of training allows a guided learning process that improves the model's parameter interpretability as certain states are associated with particular values of the labels.

\begin{figure*}[ht!]
    \centering
    \includegraphics[scale=0.47]{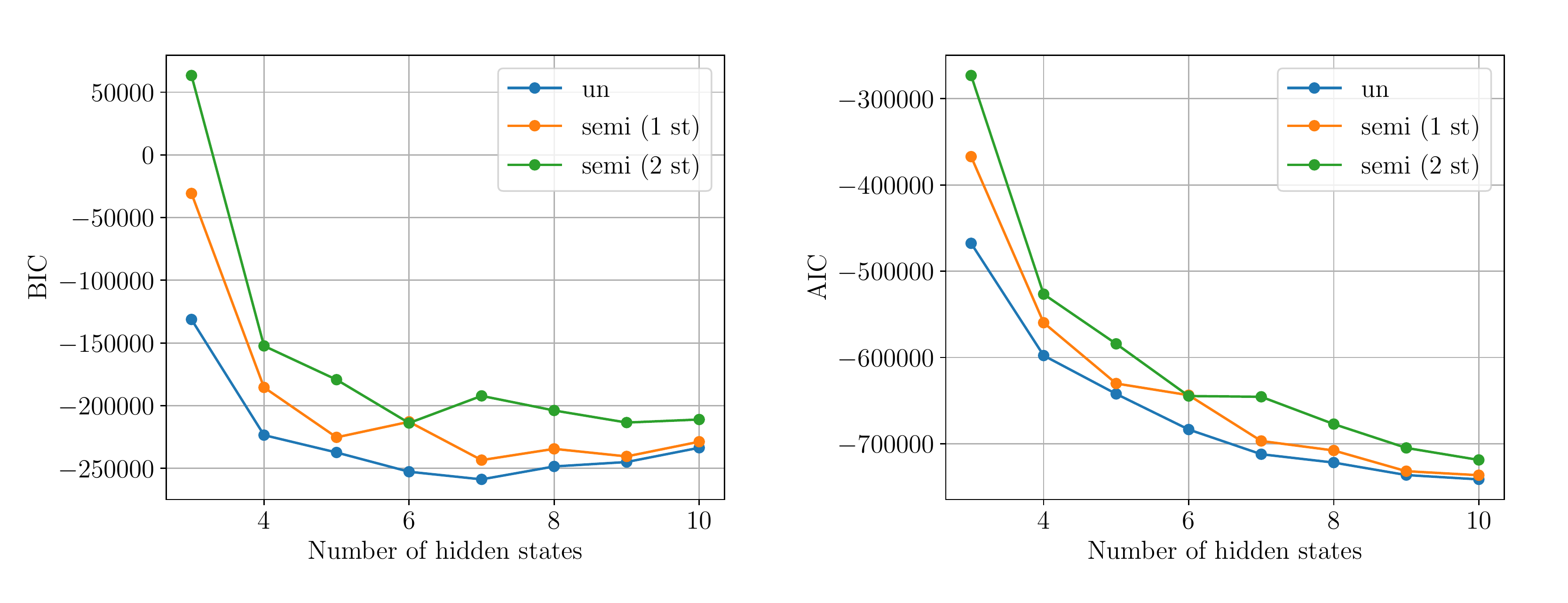}
    \caption{\ac{BIC} and \ac{AIC} obtained for the different configurations of the model: fully unsupervised (un), and semi-supervised (semi) with 1 and 2 states fixed.}\label{bic_aic}
\end{figure*}

\subsubsection{Missing Data Inference}  \label{missing_data_inference}

Missing data is relatively common in passively sensing data. These missing observations can be either entirely missing, meaning neither input signal holds a value, or partially missing, where specific signals or sources exhibit missing values. In the first case, the Baum-Welch algorithm uses the mean value for each hidden state, i.e., missing values are substituted by the average of each feature for each hidden state. In the second scenario, observations are partially missing. We infer missing values using the posterior conditional probability of the missing features given the observed ones  $p(\boldsymbol{x}_{miss}|\boldsymbol{x}_{obs}) = \mathcal{N}(\boldsymbol{x}_{missing}|\boldsymbol{\mu}_{miss|obs}, \boldsymbol{\Sigma}_{miss|obs})$ for the Gaussian features \cite{murphy2012machine}. In the case of discrete features, we use the maximum likelihood symbol for each hidden state. We refer the readers to \cite{moreno2022heterogeneous} for further details regarding the missing data inference procedure.

\subsection{Sleep Characterization}\label{sleep_characterization}

The \ac{SAR} indicator obtained through the HHMM allows us to extract informative variables that can help clinicians to portray subjects' sleep patterns. As we are not able to detect awakenings reliably due to the nature of the used features, we do not compute any related indicator as \ac{WASO} or \ac{SE}, as in order to correctly detect the awakening time the subject would have to interact with the smartphone's sensors. However, we can compute indicators related to the main sleep cycles estimated by the HHMM. Namely, we compute start and end times, duration or \ac{SPT} \cite{chaput2020sleep}, mid-sleep or \ac{CM} (midpoint between bedtime and rise-time) \cite{slavish2019intraindividual}, weekly mean start and end times, weekly mean duration or \ac{SPT}, weekly maximum and minimum \ac{SPT}, weekly mean \ac{CM}, and weekly \ac{SJ}, which is 
defined as the difference between unconstrained and constrained sleep times. We compute it as the difference between the mean \ac{CM} obtained for working days and free days, and it acts as a proxy for circadian misalignment that quantifies the discrepancy between social and biological sleep time \cite{taillard2021sleep, wittmann2006social, roenneberg2019chronotype}.


\section{Results}\label{results}

HHMM's parameters are trained via the Baum-Welch algorithm using the subset of 24 patients with 687 sequences of daily data. These training sequences are filtered to have all the signals available, less than 20\% of missing samples in actigraphy, steps, and smartphone usage signals, and less than 30\% of missing samples in the light signal. All the sequences contain data from two consecutive days, yielding vectors ranging from 14:00 to 14:00. The remainder of users are retained for model evaluation purposes, as they have wearable data available that can be used as ground truth for validation. 

The model is trained in three different configurations: completely unsupervised, semi-supervised fixing discrete emission probabilities for one state, and semi-supervised fixing discrete emission probabilities for two states. Semi-supervised training allows for a guided learning process that facilitates interpretability, as certain states are associated with particular values of the labels. In this case, we fix values of the discrete features' emission probabilities (binary step count and smartphone usage) via the emission probabilities matrix $\mathbf{B}$. For a state $q$, we fix the values of the emission probabilities so that $p(\text{steps} (t)=1|s_t=q) = 0 $ and $p(\text{smartphone} (t)=1|s_t=q) = 0 $. This way, the model will never reach state $q$ when steps or smartphone signals present a value equal to one at a given time slot $t$. We know beforehand that the detection of steps or smartphone usage is incompatible with the user being asleep. Therefore we can associate state $q$ with an asleep state. This way, we manually associate certain hidden states to low activity periods.

All the configurations are trained using a range of hidden states that varies from three to ten. Among these, the optimal number is selected using two different model selection methods: the  Bayesian Information Criterion (\ac{BIC}) and the Akaike Information Criterion (\ac{AIC}). Fig. \ref{bic_aic} shows their evolution for the different configurations and number of hidden states. Both metrics evaluate the goodness of the model's fit, but \ac{BIC} penalizes the number of free parameters more strongly.

\begin{figure}[h]%
    \centering
    \subfloat[]{{\includegraphics[scale=0.4]{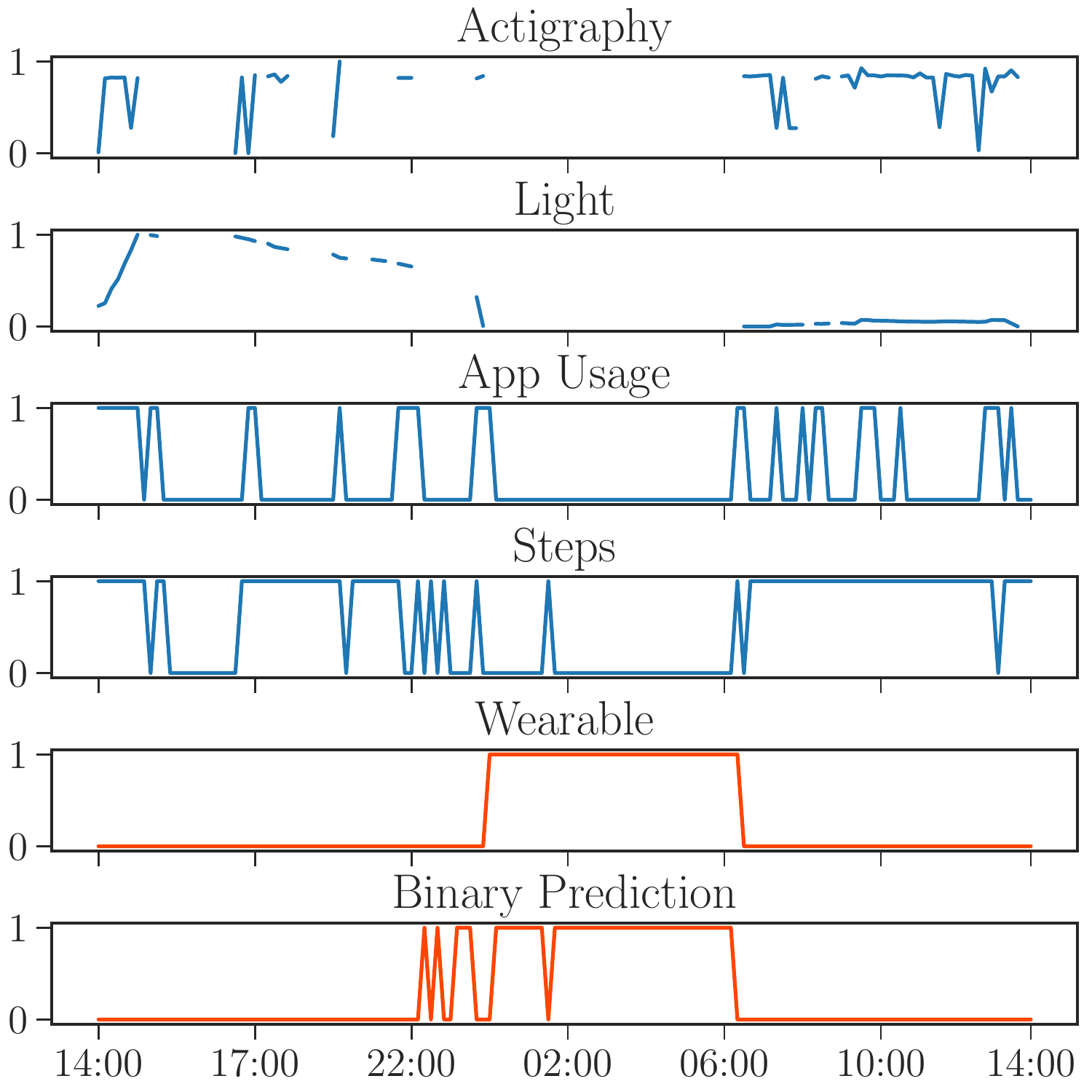} }}%
    \qquad
    \subfloat[]{{\includegraphics[scale=0.4]{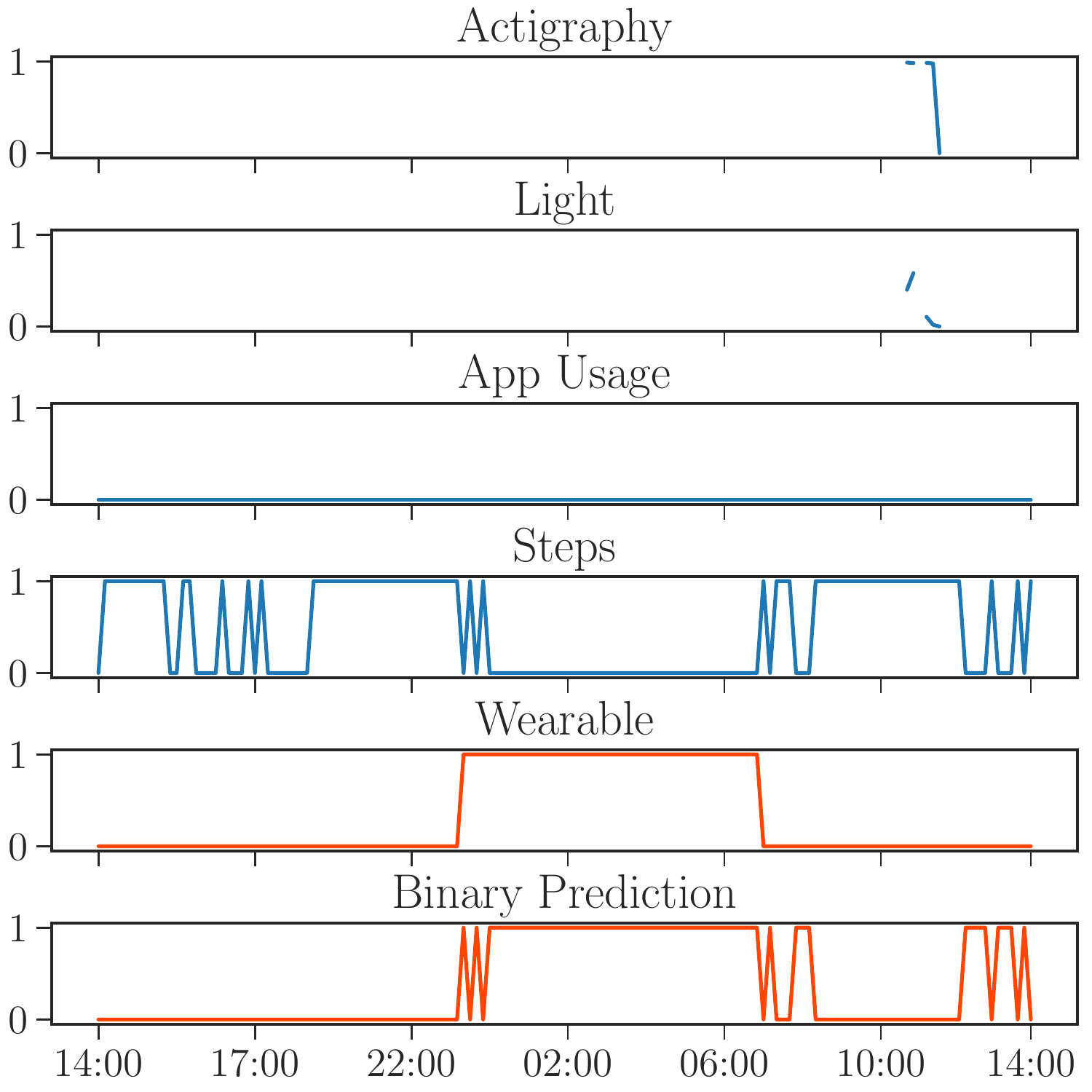} }}%
    \caption{HHMM's sleep activity predictions and wearables's reported metrics comparison. Different scenarios with varying amount of missing data were evaluated.}%
    \label{fig:example}%
\end{figure}

We select the semi-supervised configuration with six hidden states as the optimal model, fixing discrete emission probabilities for two of the six hidden states. This configuration is more interpretable and achieves a good compromise between performance and complexity, as the \ac{BIC} reflects.
Finally, predictions are obtained via the Viterbi algorithm. Mapping from hidden states to binary classification (\textit{asleep}/\textit{awake}) at each time slot is done by taking advantage of the semi-supervised configuration of the model, as we have previously mentioned. Once the algorithm finds the most likely sequence of hidden states to explain the observations, mapping from this sequence of hidden states to a binary vector is straightforward, considering we have specified the discrete probabilities of certain hidden states. Thus we have associated certain hidden states with \textit{sleep} stages. More precisely, we set to one (\textit{asleep}) the two states whose emission probabilities have been fixed, thus have been associated with sleep stages, and we set to zero (\textit{awake}) the rest of the hidden states, as shown in Fig. \ref{fig:example}. As more variability is present in the data associated with awake stages, we need a larger number of hidden states to represent these periods. Notice that many different activities take place throughout the day, and we need to associate them with an awake state.

\subsection{Sleep Activity Recognition (SAR)}\label{results_sar}

We first evaluate the model's performance as a \ac{SAR} method, i.e., detecting sleep episodes. For this purpose, we include different baseline methods for comparison. The first models considered are two dummy classifiers. The first one follows an \textit{uniform} strategy (it generates predictions uniformly at random from the list of unique classes observed in the target), referred to as \textit{Dummy (uni)} in the tables. The second dummy classifier follows a \textit{most frequent} approach (it always returns the most frequent class label observed in the target), and it is referred to as \textit{Dummy (MF)} in the tables. The second baseline considered is a K-means algorithm that assigns each data point to one of the defined clusters: \textit{asleep} or \textit{awake}. We use two variations of this model, considering different missing data imputation strategies for the binary signals. The first method, referred to as \textit{zeros} in the tables, sets to zero missing observations. The second one, referred to as \textit{MF} in the tables, substitutes missing observations with the most frequent observed value per sequence. Both methods use the empirical mean (per sequence) to infer missing values in the continuous features. Finally, we consider a \ac{GMM} that also uses the two defined imputation techniques to deal with missing entries. For all these baselines, we use scikit-learn\footnote{https://scikit-learn.org/stable/} implementations. As done with the HHMM, these baseline models are trained using the training set of 24 patients and 687 sequences.


For validation purposes, we use a set of 17 users and 1523 days of data with associated wearables' sleep metrics. These wearable devices are equipped with heart rate and oxygen saturation (SpO2) sensors, which make them accurate sleep measurement tools for nonclinical settings. While some devices only report the start and end times of the main sleep session, others report the subject's sleep as a sequence of different states: awake, light sleep, deep sleep, and REM. We transform these outputs into binary vectors that reflect two primary states: awake and asleep. By doing so, we can directly compare models' predictions and wearables' reports. Table \ref{tab:baselines1} shows the mean accuracy, sensitivity, and specificity obtained across the 1523 test sequences.

\begin{table*}[t]
\fontsize{9.0}{12}\selectfont
    \centering
    \begin{tabular}{ccccccc}
    \specialrule{.1em}{.1em}{.1em} 
                        &\multicolumn{2}{c}{\textbf{Accuracy}}   &\multicolumn{2}{c}{\textbf{Sensitivity}}    &\multicolumn{2}{c}{\textbf{Specificity}} \\
    \textbf{Model}      &\textbf{Mean}      &\textbf{STD}        &\textbf{Mean}      &\textbf{STD}            &\textbf{Mean}      &\textbf{STD}\\ \specialrule{.1em}{.1em}{.1em} 
    Dummy clf (uni)     & 0.5007            & 0.0437             & 0.5015            & 0.0766                 & 0.5004            & 0.0518 \\
    Dummy clf (MF)      & 0.6838            & 0.0540             & 0.0033            & 0.0572                 & 0.9967            & 0.0572 \\
    K-means (zeros)      & 0.5198            & 0.1650             & 0.0268            & 0.0605                 & 0.7489            & 0.2444 \\
    K-means (MF)         & 0.5195            & 0.1656             & 0.0268            & 0.0605                 & 0.7485            & 0.2454 \\
    GMM (zeros)         & 0.7275            & 0.1586             & 0.9350            & 0.1165                 & 0.6320            & 0.2259 \\
    GMM (MF)            & 0.7273            & 0.1586             & 0.9299            & 0.1303                 & 0.6341            & 0.2271 \\
    \textbf{HHMM }               & \textbf{0.8348}            & \textbf{0.1033}             & \textbf{0.8883}            & \textbf{0.1934}                 & \textbf{0.8106}            & \textbf{0.1395} \\ \specialrule{.1em}{.1em}{.1em} 
    \end{tabular}
    \caption{Results for not-filtered sequences using different models: dummy classifier following the uniform strategy (generates predictions uniformly at random from the list of unique classes observed)[Dummy clf (uni)], dummy classifier using the most frequent strategy (it returns the most frequent class label observed in the target vector)[Dummy clf (MF)], K-means imputing categorical missing values with zeros [K-means (zeros)], K-means imputing categorical missing values with the most frequent value  [K-means (MF)], GMM imputing categorical missing values with zeros [GMM (zeros)], GMM imputing categorical missing values with the most frequent value  [GMM (MF)], and HHMM.}
    \label{tab:baselines1}
\end{table*}


We should remark that this validation dataset contains sequences that present completely missing signals, as some smartphones' \ac{OS} do not allow access to specific data sources. To evaluate the performance of the models in the absence of this noise, we filter the dataset in a way that only contains sequences with all the input signals available. The resulting set contains 546 days of data from 11 different users. Table \ref{tab:baselines2} shows the results obtained for this filtered subset.

\begin{table*}[t]
\fontsize{9.0}{12}\selectfont
    \centering
    \begin{tabular}{ccccccc}
     \specialrule{.1em}{.1em}{.1em} 
                        &\multicolumn{2}{c}{\textbf{Accuracy}}   &\multicolumn{2}{c}{\textbf{Sensitivity}}    &\multicolumn{2}{c}{\textbf{Specificity}} \\
    \textbf{Model}      &\textbf{Mean}      &\textbf{STD}        &\textbf{Mean}      &\textbf{STD}            &\textbf{Mean}      &\textbf{STD}\\ \specialrule{.1em}{.1em}{.1em} 
    Dummy clf (uni)     & 0.5009            & 0.0439             & 0.4964            & 0.0756                 & 0.5002            & 0.0508 \\
    Dummy clf (MF)      & 0.6897            & 0.0452             & 0.0055            & 0.0739                 & 0.9945            & 0.0739 \\
    K-means (zeros)      & 0.5884            & 0.1124             & 0.9551            & 0.0719                 & 0.4216            & 0.1653 \\
    K-means (MF)         & 0.5889            & 0.1133             & 0.9550            & 0.0719                 & 0.4217            & 0.1665 \\
    GMM (zeros)         & 0.8058            & 0.0845             & 0.9218            & 0.1406                 & 0.7527            & 0.1042 \\
    GMM (MF)            & 0.8060            & 0.0845             & 0.9218            & 0.1406                 & 0.7531            & 0.1043 \\
    \textbf{HHMM}                & \textbf{0.8784}            & \textbf{0.0863}             & \textbf{0.8105}            & \textbf{0.2688}                 & \textbf{0.9088}            & \textbf{0.0770} \\ \specialrule{.1em}{.1em}{.1em} 
    \end{tabular}
    \caption{Results for filtered sequences using different models: dummy classifier following the uniform strategy (generates predictions uniformly at random from the list of unique classes observed) [Dummy clf (uni)], dummy classifier using the most frequent strategy (it returns the most frequent class label observed in the target vector) [Dummy clf (MF)], K-means imputing categorical missing values with zeros [K-means (zeros)], K-means imputing categorical missing values with the most frequent value  [K-means (MF)], GMM imputing categorical missing values with zeros [GMM (zeros)], GMM imputing categorical missing values with the most frequent value  [GMM (MF)], and HHMM.}
    \label{tab:baselines2}
\end{table*}


We observe that the HHMM outperforms all the baseline methods in terms of accuracy in both test datasets, being the only model that achieves good results in specificity and sensitivity simultaneously. The rest of the methods tend to overestimate one of the classes, obtaining poor results in one of these metrics. The inclusion of the dummy classifiers provides a quantification of the label distribution in the datasets, which helps in understanding class imbalances, as the portion of time the subjects are asleep is considerably smaller than the time they are awake.

In the absence of missing signals, all methods improve their metrics. In the case of the HHMM, it reaches a mean accuracy of 0.8784 across the 546 sequences. We can see that the model improves its specificity when applied to complete data, although it losses sensitivity. When all data sources are available, finding the conditions to detect an \textit{asleep} slot is more complicated. Thus the number of \textit{positives} is reduced. However, we can confirm our method is robust against missing entries, as the gain in performance removing the missing entries is just $0.0436$.

\subsection{Sleep Characterization}\label{results_sleep_characterization}

\begin{table}[h!]
\fontsize{8.0}{12}\selectfont
    \centering
    \begin{tabular}{ccccc}
     \specialrule{.1em}{.1em}{.1em} 
                        & \textbf{Mean}     & \textbf{Median}   & \textbf{Mean} & \textbf{Median}\\
    \textbf{Indicator}  & \textbf{Wearable} & \textbf{Wearable} & \textbf{HHMM} & \textbf{HHMM}\\ \specialrule{.1em}{.1em}{.1em} 
    Start               & 00:20             & 00:06             & 23:35         & 23:30\\
    End                 & 07:56             & 07:50             & 08:01         & 07:50\\
    \ac{SPT}            & 456.49 min        & 452.00 min        & 505.81 min    & 490.00 min\\
    \ac{CM}             & 04:08             & 04:01             & 03:48         & 03:45\\ \specialrule{.1em}{.1em}{.1em} 
    \end{tabular}
    \caption{Daily indicators for not-filtered sequences.}
    \label{tab:daily1}
\end{table}

\begin{table}[h!]
\fontsize{8.0}{12}\selectfont
    \centering
    \begin{tabular}{ccc}
     \specialrule{.1em}{.1em}{.1em} 
    \textbf{Metric}     & \textbf{RMSE (min)}   & \textbf{MAE (min)}\\ \specialrule{.1em}{.1em}{.1em}  
    Start               & 113.79                & 68.12 \\
    End                 & 76.81                 & 39.30 \\
    \ac{SPT}            & 119.89                & 80.30 \\
    \ac{CM}             & 76.36                 & 42.48 \\ \specialrule{.1em}{.1em}{.1em} 
    \end{tabular}
    \caption{Daily indicators' errors for not-filtered sequences.}
    \label{tab:errors_daily1}
\end{table}

\begin{table}[h!]
\fontsize{8.0}{12}\selectfont
    \centering
    \begin{tabular}{ccccc}
     \specialrule{.1em}{.1em}{.1em} 
                        & \textbf{Mean}     & \textbf{Median}   & \textbf{Mean} & \textbf{Median}\\
    \textbf{Indicator}  & \textbf{Wearable} & \textbf{Wearable} & \textbf{HHMM} & \textbf{HHMM}\\ \specialrule{.1em}{.1em}{.1em} 
    Start               & 00:19             & 00:05             & 00:15         & 00:00\\
    End                 & 07:48             & 07:42             & 07:42         & 07:40\\
    \ac{SPT}            & 448.94 min        & 447.0 min         & 447.08 min    & 450.00 min\\
    \ac{CM}             & 04:04             & 03:55             & 03:59         & 03:55\\ \specialrule{.1em}{.1em}{.1em} 
    \end{tabular}
    \caption{Daily indicators for filtered sequences.}
    \label{tab:daily2}
\end{table}

\begin{table}[h!]
\fontsize{8.0}{12}\selectfont
    \centering
    \begin{tabular}{ccc}
     \specialrule{.1em}{.1em}{.1em} 
    \textbf{Metric}     & \textbf{RMSE (min)}   & \textbf{MAE (min)}\\ \specialrule{.1em}{.1em}{.1em} 
    Start Time          & 81.84                 & 51.37 \\
    End Time            & 53.18                 & 29.45 \\
    \ac{SPT}            & 95.62                 & 64.94 \\
    \ac{CM}             & 49.77                 & 29.53 \\ \specialrule{.1em}{.1em}{.1em} 
    \end{tabular}
    \caption{Daily indicators' errors for filtered sequences.}
    \label{tab:errors_daily2}
\end{table}

Once the proposed \ac{SAR} method has been validated, we want to evaluate the quality of the information we can extract from its predictions in the form of sleep indicators, as described in Section \ref{sleep_characterization}. For this purpose, we compute the mean and median of the indicators obtained using HHMM's predictions and wearables' data, as well as the \ac{RMSE} and \ac{MAE} using wearables' reports as ground truth. Again, we obtain these metrics for the complete (17 users, 1523 days of data) and filtered (11 users, 546 days of data) test datasets. Tables \ref{tab:daily1},  \ref{tab:daily2}, \ref{tab:errors_daily1} and \ref{tab:errors_daily2} contain the results relative to daily indicators, while tables \ref{tab:weekly1}, \ref{tab:weekly2}, \ref{tab:errors_weekly1}, and \ref{tab:errors_weekly2} contain the results relative to weekly indicators, as the \ac{SJ}.

In the case of weekly indicators, as the \ac{SJ} compares the mean CM obtained for working and free days, we need to consider those weeks containing at least four days of data with both weekend and workday records. This first filtering yields a subset of data with 995 weeks and 17 different users. Results obtained for the analysis of this dataset are contained in Tables \ref{tab:weekly1} and \ref{tab:errors_weekly1}. Then, we filter this dataset again to isolate complete observations (all sources available), resulting in a data subset of 201 weeks and 11 different users. Results obtained for the analysis of this second dataset are contained in Tables \ref{tab:weekly2} and \ref{tab:errors_weekly2}.

\begin{table}[ht]
\fontsize{8.0}{12}\selectfont
    \centering
    \begin{tabular}{ccccc}
     \specialrule{.1em}{.1em}{.1em} 
                        & \textbf{Mean}     & \textbf{Median}   & \textbf{Mean} & \textbf{Median}\\
    \textbf{Indicator}  & \textbf{Wearable} & \textbf{Wearable} & \textbf{HHMM} & \textbf{HHMM}\\ \specialrule{.1em}{.1em}{.1em} 
    Mean Start          & 00:05             & 00:05             & 23:27         & 23:25\\
    Mean End            & 07:42             & 07:50             & 07:51         & 07:53\\
    Mean \ac{SPT}       & 457.02 min        & 455.91 min        & 503.65 min    & 508.00 min\\
    Max \ac{SPT}        & 534.30 min        & 531.50 min        & 619.83 min    & 610.00 min\\
    Min \ac{SPT}        & 387.41 min        & 399.00 min        & 408.05 min    & 410.00 min\\
    Mean \ac{CM}        & 03:54             & 03:59             & 03:39         & 03:47\\
    \ac{SJ}             & 73.31 min         & 63.00 min         & 68.26 min     & 64.0 min\\ \specialrule{.1em}{.1em}{.1em} 
    \end{tabular}
    \caption{Weekly indicators for not-filtered sequences.}
    \label{tab:weekly1}
\end{table}

\begin{table}[ht]
\fontsize{8.0}{12}\selectfont
    \centering
    \begin{tabular}{ccc}
     \specialrule{.1em}{.1em}{.1em} 
    \textbf{Metric}     & \textbf{RMSE (min)}   & \textbf{MAE (min)}\\ \specialrule{.1em}{.1em}{.1em} 
    Mean Start          & 70.57                 & 52.16 \\
    Mean End            & 41.67                 & 27.21 \\
    Mean \ac{SPT}       & 79.28                 & 63.83 \\
    Max \ac{SPT}        & 129.76                & 97.54 \\
    Min \ac{SPT}        & 91.57                 & 60.12 \\
    Mean \ac{CM}        & 42.28                 & 26.01 \\
    \ac{SJ}             & 95.42                 & 38.91 \\ \specialrule{.1em}{.1em}{.1em} 
    \end{tabular}
    \caption{Weekly indicators' errors for not-filtered sequences.}
    \label{tab:errors_weekly1}
\end{table}

\begin{table}[ht]
\fontsize{8.0}{12}\selectfont
    \centering
    \begin{tabular}{ccccc}
     \specialrule{.1em}{.1em}{.1em} 
                        & \textbf{Mean}     & \textbf{Median}   & \textbf{Mean} & \textbf{Median}\\
    \textbf{Indicator}  & \textbf{Wearable} & \textbf{Wearable} & \textbf{HHMM} & \textbf{HHMM}\\ \specialrule{.1em}{.1em}{.1em} 
    Mean Start          & 00:05             & 00:00             & 00:16         & 23:51\\
    Mean End            & 07:29             & 07:31             & 07:20         & 07:08\\
    Mean \ac{SPT}       & 444.15 min        & 442.08 min        & 424.49 min    & 418.00 min\\
    Max \ac{SPT}        & 508.16 min        & 508.50 min        & 522.73 min    & 510.00 min\\
    Min \ac{SPT}        & 393.52 min        & 397.00 min        & 335.57 min    & 340.00 min\\
    Mean \ac{CM}        & 03:47             & 03:48             & 03:48         & 03:56\\
    \ac{SJ}             & 69.76 min         & 69.00 min         & 66.92 min     & 64.00 min\\ \specialrule{.1em}{.1em}{.1em} 
    \end{tabular}
    \caption{Weekly indicators for filtered sequences.}
    \label{tab:weekly2}
\end{table}

\begin{table}[ht]
\fontsize{8.0}{12}\selectfont
    \centering
    \begin{tabular}{ccc}
     \specialrule{.1em}{.1em}{.1em} 
    \textbf{Metric}     & \textbf{RMSE (min)}   & \textbf{MAE (min)}\\ \specialrule{.1em}{.1em}{.1em} 
    Mean Start          & 47.09                 & 35.91 \\
    Mean End            & 31.68                 & 18.52 \\
    Mean \ac{SPT}       & 61.59                 & 45.72 \\
    Max \ac{SPT}        & 75.70                 & 47.29 \\
    Min \ac{SPT}        & 110.42                & 77.25 \\
    Mean \ac{CM}        & 25.73                 & 16.90 \\
    \ac{SJ}             & 33.55                 & 22.77 \\ \specialrule{.1em}{.1em}{.1em} 
    \end{tabular}
    \caption{Weekly indicators' errors for filtered sequences.}
    \label{tab:errors_weekly2}
\end{table}

In light of the results, we can affirm that the model presents a more significant error in the start time estimation (for both daily and weekly indicators) than in the end time estimation. This phenomenon can be due to the lack of measurements once the subject is lying in bed. As the proposed method relies on smartphone passively sensed data and user-smartphone interaction, it is impossible to measure the \ac{SOL}, which is the time it takes to transition from full wakefulness to sleep. In the same way, as it is a common habit to check the phone first thing in the morning (to turn off the alarm, check the hour, check notifications, etc.), we can detect end sleep times more accurately. Wearable devices do not face this limitation, and they can provide more accurate estimations because they are equipped with additional sensors that can measure users' vital signs, such as breathing rate or heartbeat, being able to differentiate whether the subject is lying in bed awake or asleep.

The errors contained in the estimation of the start and end times propagate in the computation of the remainder indicators: \ac{SPT}, \ac{CM}, and \ac{SJ}, increasing the overall error of our sleep characterization. This could be partially solved if we could estimate the \ac{SOL} for each user. Although it is impossible to obtain such a measure utilizing our defined data sources, it could be assessed by external methods such as initial questionnaires to the subjects.

\section{Discussion}

In this work, we have introduced a general \ac{SAR} method to estimate and characterize sleep based on passively sensed data from smartphones. We have integrated this data into a single behavior metric utilizing a temporal discrete latent variable model: the \ac{HHMM}. The proposed method can successfully handle problems and limitations usually associated with passively sensed data, such as data heterogeneity, corrupted signals, and missing values while achieving reliable results. We have compared the performance of our model against different state-of-the-art methods, obtaining better performance in detecting sleep episodes. Additionally, we have validated the performance of our method against wearables’ reported sleep metrics for both sleep estimation and characterization, achieving good precision and thus proving the effectiveness of the proposed approach, which could be used as a general method to assess sleep when no wearable device is available.


The main limitation of the proposed method is the inability to distinguish sleep sessions from periods with no reported activity, such as when the device is charging, for example. For this reason, the model cannot differentiate time in bed from time asleep, which introduces a significant error in estimating sleep start time, propagated in the computation of the remainder indicators. Although this limitation cannot be solved from the model's point of view (it arises from the data source itself), it could be alleviated by using personalized post-processing. Introducing additional information about the subjects, such as work
schedules or SOL estimation, which could be assessed utilizing initial questionnaires, the model's predictions could be adjusted for each subject in a personalized manner.

Given the continuous and non-invasive nature of the proposed method, it could be used for patient behavior monitoring. For this specific purpose, it is not necessary to obtain precise indicators but coherent markers over time that can exhibit a limited amount of error. Since the main limitation of our model is directly related to the data sources, errors in the predictions should be stable as long as the users' habits remain stable. Significant changes detected in the estimated sleep indicators can be directly linked to changes in their routine and habits. Therefore, the indicators estimated by our method can constitute an informative source for \ac{CPD} algorithms \cite{moreno2021change} to study subjects' habits and routines. 

In conclusion, we have proved that the proposed method is a feasible option to monitor and characterize sleep when a more reliable source, such as wearable devices, is unavailable. It provides a good estimation of sleep indicators that can be used to define behavioral patterns, and monitoring evolution and changes in subjects' sleep routines and daily habits.

\section{Acknowledgements}

This work has been partly supported by the Spanish Ministerio de Ciencia, Innovación y Universidades (FPU18/00470), Ministerio de Ciencia e Innovación jointly with the European Commission (ERDF) (PID2021-123182OB-I00, PID2021-125159NB-I00), by Comunidad de Madrid (IntCARE-CM and IND2020/TIC-17372), and by the European Union (FEDER) and the European Research Council (ERC) through the European Union’s Horizon 2020 research and innovation program under Grant 714161.

\bibliographystyle{IEEEbib}
\bibliography{mlsp_2020}

\end{document}